\documentclass[preprint]{elsarticle}

\usepackage{hyperref,amsmath,graphicx}
\biboptions{sort&compress}

\journal{Journal of Magnetism and Magnetic Materials}

\begin{document}

\begin{frontmatter}

\title{Thermodynamic features of the 1D dilute Ising model in the external magnetic field}
\author[1]{A.V. Shadrin\corref{cor1}}
\ead{shadrin.anton@urfu.ru}
\author[1]{Yu.D. Panov}
\address[1]{Institute of Natural Sciences and Mathematics, Ural Federal University, 620002, 19 Mira street, Ekaterinburg, Russia}
\cortext[cor1]{Corresponding author}

\begin{abstract}
We consider the effects of the magnetic field on the frustrated phase states of the dilute Ising chain, especially,  
the behavior of the magnetic entropy change and the isentropic dependence of the temperature on the magnetic field, which are the key parameters of the magnetocaloric effect.
The found temperature dependences of entropy demonstrate the nonequivalence of frustrated phases in the antiferromagnetic and ferromagnetic cases.
In the antiferromagnetic case, the nonzero magnetic field at certain parameters causes a charge ordering for nonmagnetic impurities at a half-filling, while in the ferromagnetic case, the magnetic field reduces the frustration of the ground state only partially. 
It is also shown, that impurities radically change the magnetic Gr\"uneisen parameter in comparison with the case of a pure Ising chain.
\end{abstract}

\begin{keyword}
dilute Ising chain \sep 
frustrated magnets \sep
magnetic entropy change
\end{keyword}

\end{frontmatter}

\section{Introduction}

One of the remarkable features of low-dimensional systems, such as 
decorated Ising models~\cite{Canova2006,Torrico2014,Strecka2020,Galisova2015,Rojas2016,Strecka2016,DeSouza2018}, 
the anisotropic Potts chain~\cite{Panov2021}, 
the diamond Hubbard chain~\cite{Rojas2021}, 
is the presence, under certain parameters, of a frustrated ground state for which the residual entropy is nonzero. 
Despite the absence of a real phase transition at finite temperatures according to the Perron--Frobenius theorem for square real matrices~\cite{Ninio1976} the thermodynamic behavior near the boundaries between different phases of the ground state for these systems can exhibit striking features. 
As shown in~\cite{Rojas2020}, if one of the phases has a nonzero residual entropy that preserves continuity at the boundary with the other phase, then the thermodynamic characteristics of the system will demonstrate pseudo-transitions at a finite temperature. 
Entropy, heat capacity, magnetization, and susceptibility have similar features to the behavior of these properties at conventional phase transitions, including the presence of quasicritical exponents~\cite{Rojas2019}.

Recently, frustrated magnetic systems have also attracted the attention of researchers due to the enhanced magnetocaloric effect in the vicinity of finite-field transitions~\cite{Zhitomirsky2003,Zhitomirsky2004}. 
Besides to geometric factors, the impurities are also the reason for the existence of frustrations in the magnetic system. 
The simplest example of a magnetic system that is frustrated by impurities is a diluted Ising chain. 
The Hamiltonian of 1D diluted Ising model can be written in the following form
\begin{equation}
	H = -J \sum_{i} S_{z,i} S_{z,i+1} 
	+ V \sum_{i} P_{0,i} P_{0,i+1} - h \sum_{i} S_{z,i} 
	- \mu \sum_{i} P_{0,i} .
	\label{eq:Ham}
\end{equation}
Here we use the $S=1$ pseudospin operators. 
The states for a given lattice site with the pseudospin projections $S_z = \pm 1$ correspond to the two magnetic states with the conventional spin projections $s_z = \pm 1/2$, while the state with $S_z = 0$ corresponds to the charged nonmagnetic state.
$S_{z,i}$ is a $z$-projection of the on-site pseudospin operator, 
$P_{0,i} = 1 - S_{z,i}^2$ is the projection operator onto the $S_z=0$ state, 
$J$ is the exchange constant, 
$V>0$ is the inter-site correlation parameter for impurities,  
$h$ is an external magnetic field, 
and $\mu$ is a chemical potential for impurities. 
Further we will assume that nonmagnetic impurities are mobile, which corresponds to the annealed system. 
As well known~\cite{Hill1956}, 
$V = V_0 + V_1 -2 V_{01}$ describes the interaction for a more general case:
\begin{equation}
	V_0 \sum_{i} P_{0,i} P_{0,i+1} 
	+ V_1 \sum_{i} P_{1,i} P_{1,i+1} 
	+ V_{01} \sum_{i} \Big( P_{0,i} P_{1,i+1} + P_{1,i} P_{0,i+1} \Big)
	,
\end{equation}
where $P_{1} = S_{z,}^2$ is the projection operator onto magnetic states.
The solutions and various thermodynamic properties of the 1D dilute Ising model at zero external magnetic field was found in~\cite{Rys1969,Matsubara1973,Termonia1974,Balagurov1974,Panov2020}.
If $h\neq0$, then there are no explicit analytical expressions for various thermodynamic functions of the model~\eqref{eq:Ham}. 
It is known the account of magnetic field for the $S=1$ Ising chain 
significantly expands the list of possible phase states of the system and leads to various features of thermodynamic behavior~\cite{Kassan-ogly2001,Zarubin2019}.

In the present paper, we consider the effects of the magnetic field on the frustrated phase states of the model~\eqref{eq:Ham}. 
We focused on the behavior of entropy and, in particular, on the magnetic entropy change, which is the key parameter of the magnetocaloric effect. 
Also, we explore the isentropic dependence of the temperature on the magnetic field. 
The paper is organized as follows. 
We briefly describe the methods in section 2, and section 3 the results, 
including the ground state phase diagram, and their discussions are given. 
Conclusions are presented in section 4.

\section{Methods}
We define the transfer matrix for the model~\eqref{eq:Ham} as
\begin{equation}
\tau = \left(
\begin{array}{cccc}
xz & z^{1/2} \, t^{1/2} & x^{-1} \\
z^{1/2} \, t^{1/2} & y^{-1} t & z^{-1/2} \, t^{1/2} \\
x^{-1} & z^{-1/2} \, t^{1/2} & x z^{-1} 
\end{array}
\right),
\label{eq:Tr-matrix}
\end{equation}
where $x= e^{\beta J}$, $y = e^{\beta V}$, $z = e^{\beta h}$, $t = e^{\beta \mu}$ and  $\beta = 1/T$, and we assume $k_B = 1$. From \eqref{eq:Tr-matrix}, we found the characteristic equation for the eigenvalues $\lambda_i$:
\begin{multline}\label{CharacterEq}
	\lambda^3 - \lambda^2 \left( ty^{-1} + x ( z + z^{-1} ) \right) 
		- \lambda \left( x^2 - x^{-2} + t \left( xy^{-1} - 1 \right) ( z + z^{-1} ) \right) \\
	{}- 2t (x - x^{-1}) - t x^{-2} y^{-1} = 0.
\end{multline}
The eigenvalues in a general case are cumbersome functions, but at $h=0$ they could be reduced to the known expressions~\cite{Panov2020}:
\begin{eqnarray}\label{Eigenvalues_h=0}
	\lambda_{1,2} & = & 
	\frac{1}{2} \left(x + x^{-1} + y^{-1} t\right) 
	\pm \left[ 2t + \frac{1}{4} \left( x + x^{-1} - y^{-1} t \right)^2 \right]^{1/2}
	,	\nonumber \\
	\lambda_3 & = & x - x^{-1} .
\end{eqnarray}
According to the Perron--Frobenius theorem~\cite{Ninio1976}, there is only one maximum eigenvalue, $\lambda_1$, and in the thermodynamic limit we obtain the grand potential  and the entropy in the following form:
\begin{equation}
	\Omega = N \omega = - N T \ln \lambda_1 ,\qquad
	S = - \left(\frac{\partial \omega}{\partial T}\right)_{h,\mu}  
	= \ln \lambda_1 + \frac{T}{\lambda_1} \left(\frac{\partial \lambda_1}{\partial T}\right)_{h,\mu} .
	\label{eq:OmS}
\end{equation}
The  grand potential and entropy found depend on parameters $J$, $V$, $h$, $\mu$, and $T$. 
But in the present problem,  it is more convenient to use the concentration $n$ of impurities as an external parameter.
The dependence $n(\mu)$ can be obtained from the equation
\begin{equation}
	n = - \left(\frac{\partial \omega}{\partial \mu}\right)_{T,h} 
	= \frac{T}{\lambda_1} \left(\frac{\partial \lambda_1}{\partial \mu}\right)_{T,h} .
	\label{eq:mu}
\end{equation}

In a general case $h\neq0$, we used numerical methods to get the inverse dependence $\mu(n)$  
and fix the concentration of impurities $n$ at all temperatures.
If $h=0$, we obtain the explicit expressions~\cite{Panov2020}:
\begin{equation}
	\mu = \ln \left[ y \left( x + x^{-1} \right) \frac{ g + m  }{  g - m  } \right] , 
\end{equation}
\begin{multline}
	S = 
	\frac{1}{2}  \ln \frac{ 2 \left( 1+2g \right)^2 }{ 1 - 4m^2 }
	+ \frac{g + 2m^2}{1+2g} \, \ln y
	- m \ln \left[ \left( x + x^{-1} \right) \frac{ g + m  }{  g - m  } \right]
	\\
	{}- \left( 1-2m \right) 
	\frac{\left(g - m\right)\left(x - x^{-1}\right)}{\left(1+2g\right)\left(x + x^{-1}\right)} \, \ln x ,
	\label{eq:s}
\end{multline}
where
\begin{equation}
	g = \left[ m^2 + \frac{1}{2}\left( \frac{1}{4} - m^2 \right) y^{-1} \left( x + x^{-1} \right) \right]^{1/2} ,
	\label{eq:g}
\end{equation}
and we introduced the deviation of the concentration of impurities from half-filling, $m = n-1/2$.

The knowledge of the entropy from Eqs.~(\ref{eq:OmS},\ref{eq:mu}) gives an opportunity to explore magnetocaloric properties of the dilute Ising chain for a given $n$. 
We explore the magnetic entropy change, 
the isentropic dependencies of the temperature on the magnetic field 
and the magnetic Gr\"uneisen parameter, which can be calculated from the relation
\begin{equation}
	\Gamma_{mag} = \frac{1}{T} \left( \frac{\partial T}{\partial h} \right)_{S,n} 
	= - \frac{1}{T} 
	\frac{\left( \partial S / \partial h \right)_{T,n}}{\left( \partial S / \partial T \right)_{h,n}}.
	\label{eq:GrunDef}
\end{equation}
The explicit expression that we use to calculate $\Gamma_{mag}$ for a given $n$ 
in variables $(T,h,\mu)$ has the following form:
\begin{equation}
	\Gamma_{mag} = - \frac{1}{T} 
	\left[
	\frac{
	\left( {\partial S / \partial h} \right)_{T,\mu}
	\left( {\partial n / \partial \mu} \right)_{T,h}
	- \left( {\partial S / \partial \mu} \right)_{T,h}
	\left( {\partial n / \partial h} \right)_{T,\mu}
	}{
	\left( {\partial S / \partial T} \right)_{h,\mu}
	\left( {\partial n / \partial \mu} \right)_{T,h}
	- \left( {\partial S / \partial \mu} \right)_{T,h}
	\left( {\partial n / \partial T} \right)_{h,\mu}
	}
	\right].
\end{equation}

\section{Results}

\subsection{Phase diagram at zero temperature.}

\begin{figure}
\centering
	\includegraphics[width=0.48\textwidth]{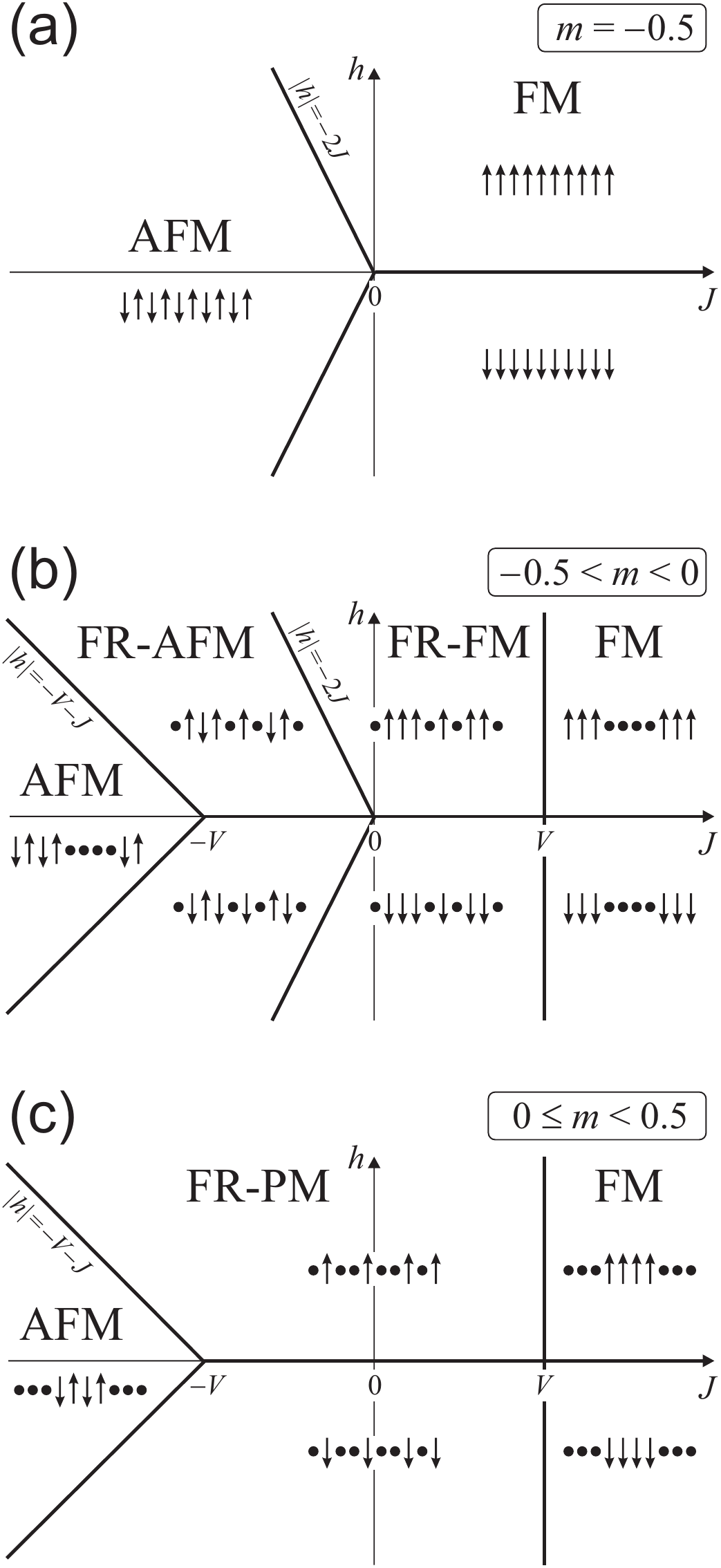}
	\caption{
	Phase diagram at zero temperature for a dilute Ising chain in a longitudinal magnetic field: 
	(a) the Ising chain without impurities, $n=0$,
	(b) the case of a weakly diluted chain,
	(c) the case of a strongly diluted chain.
	\label{fig:phdiag}
	}
\end{figure}

The phase diagram of the dilute Ising chain in longitudinal magnetic field at zero temperature is shown in Fig.~\ref{fig:phdiag} for the $J-h$ plane. 
The limiting case for the Ising chain without impurities, $m=-1/2$, 
is given in Fig.~\ref{fig:phdiag}(a). 
Two ground states, the ferromagnetic (FM) state with magnetic moment oriented towards the field, 
and the antiferromagnetic (AFM) state with zero magnetic moment, 
are separated by the critical value of magnetic field $|h_c| = -2J$ 
at which the spin-flip transition occurs. 
Figures~\ref{fig:phdiag}(b) and ~\ref{fig:phdiag}(c) 
show the cases of a weakly diluted chain, $-1/2<m<0$, 
and a strongly diluted chain, $0\leq m<1/2$, respectively. 
Dilution with impurities leads to the appearance of two new boundary lines on the $J-h$ plane, 
$J=V$ and $|h|=-J-V$. 
If $J>V$, the ground state is represented by macroscopic FM domains (or drops) separated by macroscopic impurity domains. 
Similarly, the AFM domains arise when $J<-V-|h|$. 
Schematically, this is shown in Fig.~\ref{fig:phdiag}(b,c), where the arrows correspond to the spins, and the circles correspond to the impurities. 
For both FM and AFM phases, the entropy is zero. 

Analysis of the ground state of the model~\eqref{eq:Ham} in zero magnetic field shows~\cite{Balagurov1974,Panov2020} that phases with the nonzero residual entropy exist at $|J|<V$ and at $|J|=V$. 
For the weak exchange, when $|J|<V$, the spin correlation length is always finite, 
but the impurity correlation length with the temperature lowering tends to infinity at the half-filling concentration, $m=0$, due to the formation of charge ordering~\cite{Panov2020}. 
If $|J|=V$ and $h=0$, the spin correlation length and the impurity correlation length are finite for all values of the impurity concentration and temperature.

In magnetic field, if $-V-|h|<J<V$, the residual entropy is also not zero, so the ground state is frustrated. 
If $-1/2<m<0$, the ground state of the chain is a set of finite AFM or FM spin clusters separated by impurities. 
This state we call frustrated ferromagnetic (FR-FM) or frustrated antiferromagnetic (FR-AFM) respectively. 
The FR-FM and FR-AFM states separated by the spin-flip line, $|h| = -2J$, 
as it is shown in Fig.~\ref{fig:phdiag}(b). 
In the strongly diluted case, $0\leq m<1/2$, there are single spins separated by impurity clusters, and the system exhibits a paramagnetic response, which is uniform over the entire range 
$-V-|h|<J<V$. 
This frustrated paramagnetic (FR-PM) state is shown in Fig.~\ref{fig:phdiag}(c).

\subsection{The magnetic entropy change.}

\begin{figure}
\centering
	\includegraphics[width=\textwidth]{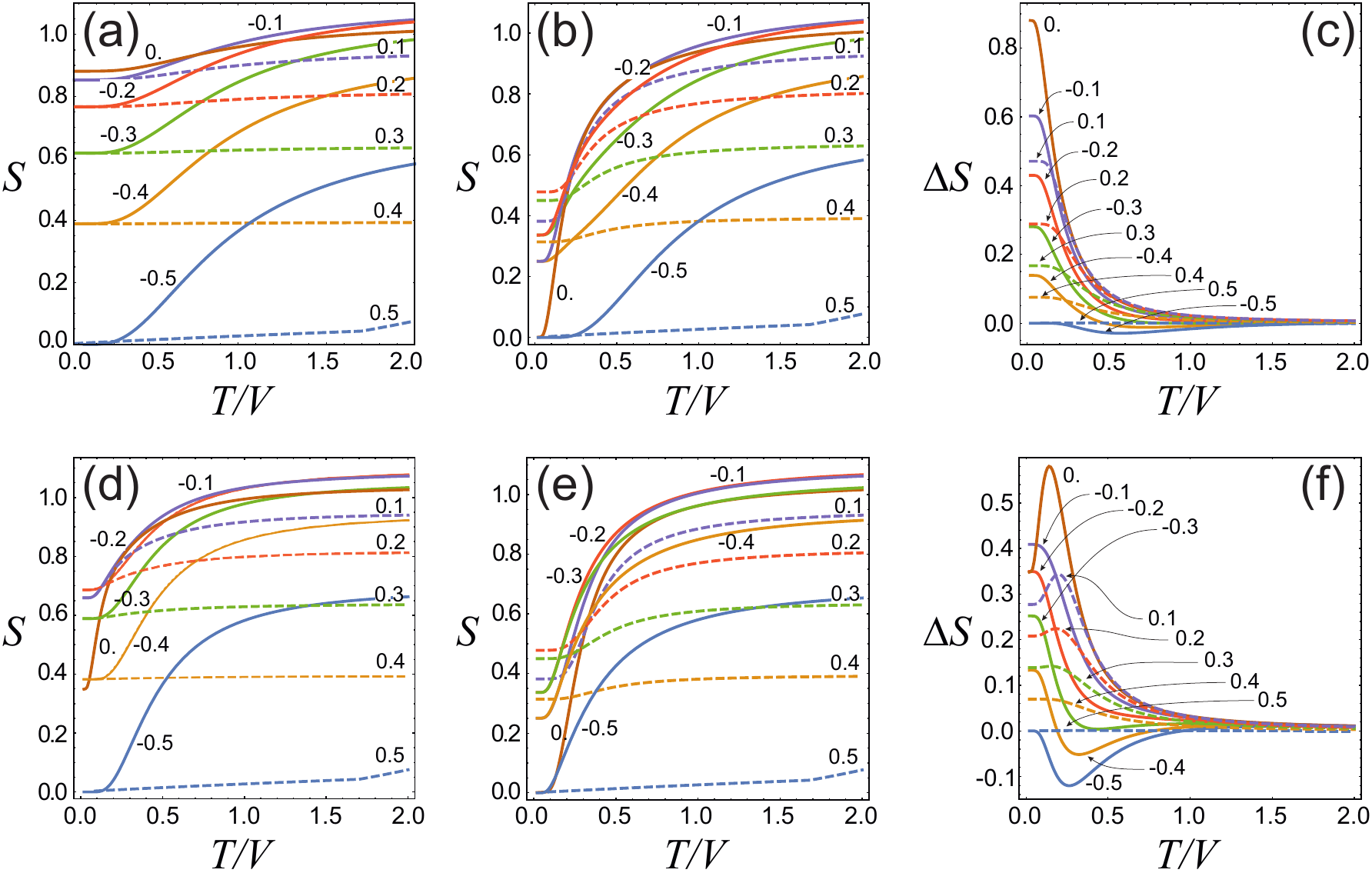}
	\caption{(color online) 
	Temperature dependences of the entropy $S$ and the magnetic entropy change $\Delta S$ in the AFM case ($J<0$). 
	Panels (a), (b), and (c) correspond to $J/V=-1$; (d), (e), (f) -- to $J/V = -0.5$. 
	Panels (a) and (d) show the entropy $S$ at $h=0$, (b) and (e) -- at $h/V=0.5$, 
	(c) and (f) -- the magnetic entropy change  $\Delta S = S(h=0)-S(h=0.5)$. 
	The numbers near lines correspond to the deviation of the impurity concentration $n$ from half-filling, $m = n-1/2$. 
	Solid (dashed) lines correspond to $m\leq0$ ($m>0$).
	\label{fig:negJ}
	}
\end{figure}

\begin{figure}
\centering
	\includegraphics[width=\textwidth]{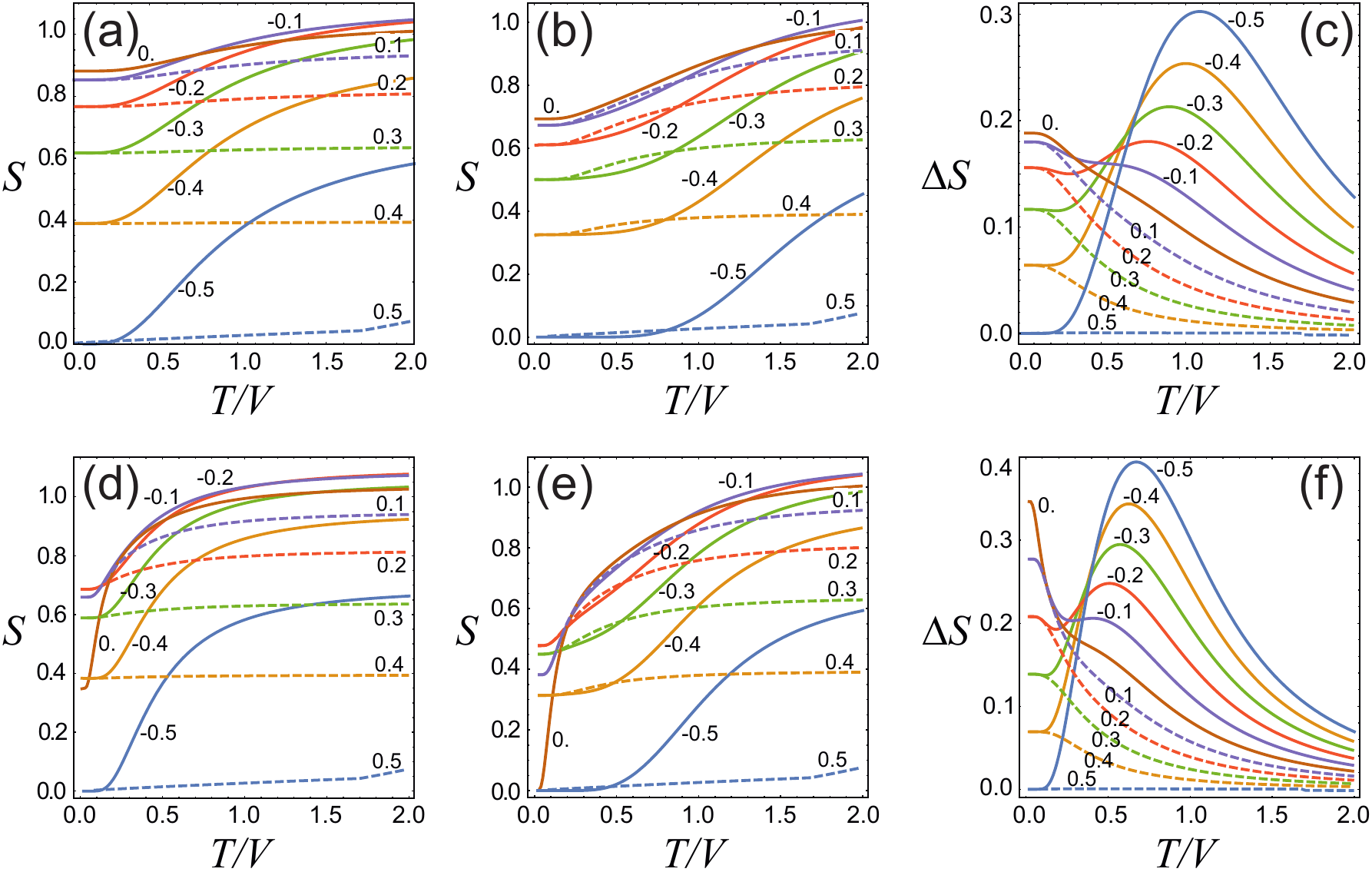}
	\caption{(color online)  
	Temperature dependences of the entropy $S$ and the magnetic entropy change $\Delta S$ in the FM case ($J>0$). 
	Panels (a), (b), and (c) correspond to $J/V=1$; (d), (e), (f) -- to $J/V = 0.5$. 
	Panels (a) and (d) show the entropy $S$ at $h=0$, (b) and (e) -- at $h/V=0.5$, 
	(c) and (f) -- the magnetic entropy change $\Delta S = S(h=0)-S(h=0.5)$. 
	The numbers near lines correspond to the deviation of the impurity concentration $n$ from half-filling, $m = n-1/2$. 
	Solid (dashed) lines correspond to $m\leq0$ ($m>0$).
	\label{fig:posJ}
	}
\end{figure}

Temperature dependences of the entropy $S$ and the magnetic entropy change, $\Delta S = S(h=0)-S(h\neq0)$, are shown 
in Fig.~\ref{fig:negJ} for the antiferromagnetic (AFM) sign of the exchange constant, $J<0$, 
and in Fig.~\ref{fig:posJ} for the ferromagnetic (FM) sign, $J>0$. 
The correlation parameter for impurities $V$ accepted and used as a positive scaling factor.

Fig.~\ref{fig:negJ} shows the temperature dependences of the entropy for $J/V=-1$, $h=0$ in panel (a), 
and for  $J/V=-1$, $h/V=0.5$ in panel (b). 
If at $h=0$ the entropy monotonically depends on $|m|$ and has a maximum at $m=0$, 
at $h\neq0$ the dependence on $|m|$ has a local minimum at $m=0$.
The magnetic entropy change for $J/V=-1$ is shown in panel (c). 
The maximum $\Delta S$ for all $m$ is achieved at $T=0$
and also has a minimum with $\Delta S<0$ at finite temperature for small values of impurity concentrations.

It is worth noting that in the AFM chain, for any value of the applied magnetic field, we get zero entropy at $m=0$, because there is only one way to minimize the energy: alternating spins and charges, and all spins are oriented along the magnetic field. 
In a certain sense, in this case we get a kind of magneto-electric effect: an external magnetic field causes a charge ordering. 
This also gives us the maximum change in entropy at half-filling.

The temperature dependences of the entropy for $J/V=-0,5$, $h=0$ are shown in Fig.~\ref{fig:negJ}(d), 
and for  $J/V=-0.5$, $h/V=0.5$ in Fig.~\ref{fig:negJ}(e). 
The dependences of $S$ on $|m|$ have a local minimum at $m=0$ both at $h=0$ and at $h/V=0.5$.
The magnetic entropy change for $J/V=-0.5$ is shown in Fig.~\ref{fig:negJ}(f). 
In a contrast to the previous case, the $\Delta S$ dependences show a maximum at finite temperature for some $m\geq0$, 
and also show a minimum with $\Delta S<0$ at finite temperature in some range for $m<0$.

Fig.~\ref{fig:posJ} shows the temperature dependences of the entropy for $J/V=1$, $h=0$ in panel (a), 
and for  $J/V=1$, $h/V=0.5$ in panel (b). 
The magnetic entropy change for $J/V=1$ is shown in panel (c). 
Both at $h=0$ and $h\neq0$ the entropy monotonically depends on $|m|$ and has a maximum at $m=0$.
The magnetic entropy change has a maximum at finite temperature for some $m<0$, and near the $m=0$ 
it also has a local minimum at a finite temperature. 
In the case of FM, we will not get the same effect at $h > 0$ as for $J/V=-1$, because it makes no sense to split the spin clusters into more than one spin in order to minimize the energy. 
But the entropy is still slightly reduced, because there is no ordering chaos for different the spin clusters: they will all be oriented by a magnetic field.

The temperature dependences of the entropy for $J/V=0.5$, $h=0$ are shown Fig.~\ref{fig:posJ}  in panel (d), 
and for  $J/V=0.5$, $h/V=0.5$ in panel (e). 
The magnetic entropy change for $J/V=0.5$ is shown in panel (f). 
Qualitatively, the behavior of entropy differs from $J/V=1$ case at some region near $|m|=0$, where the tendency to the charge ordering causes the decreasing of $S$. 
The $\Delta S$ dependences also show local maxima at finite temperature in some range for $m<0$, 
and a monotonic behavior with maximal value at $T=0$ for $m\geq0$. 
The magnetic entropy change for FM case is always positive.

\begin{figure}
\centering
	\includegraphics[width=\textwidth]{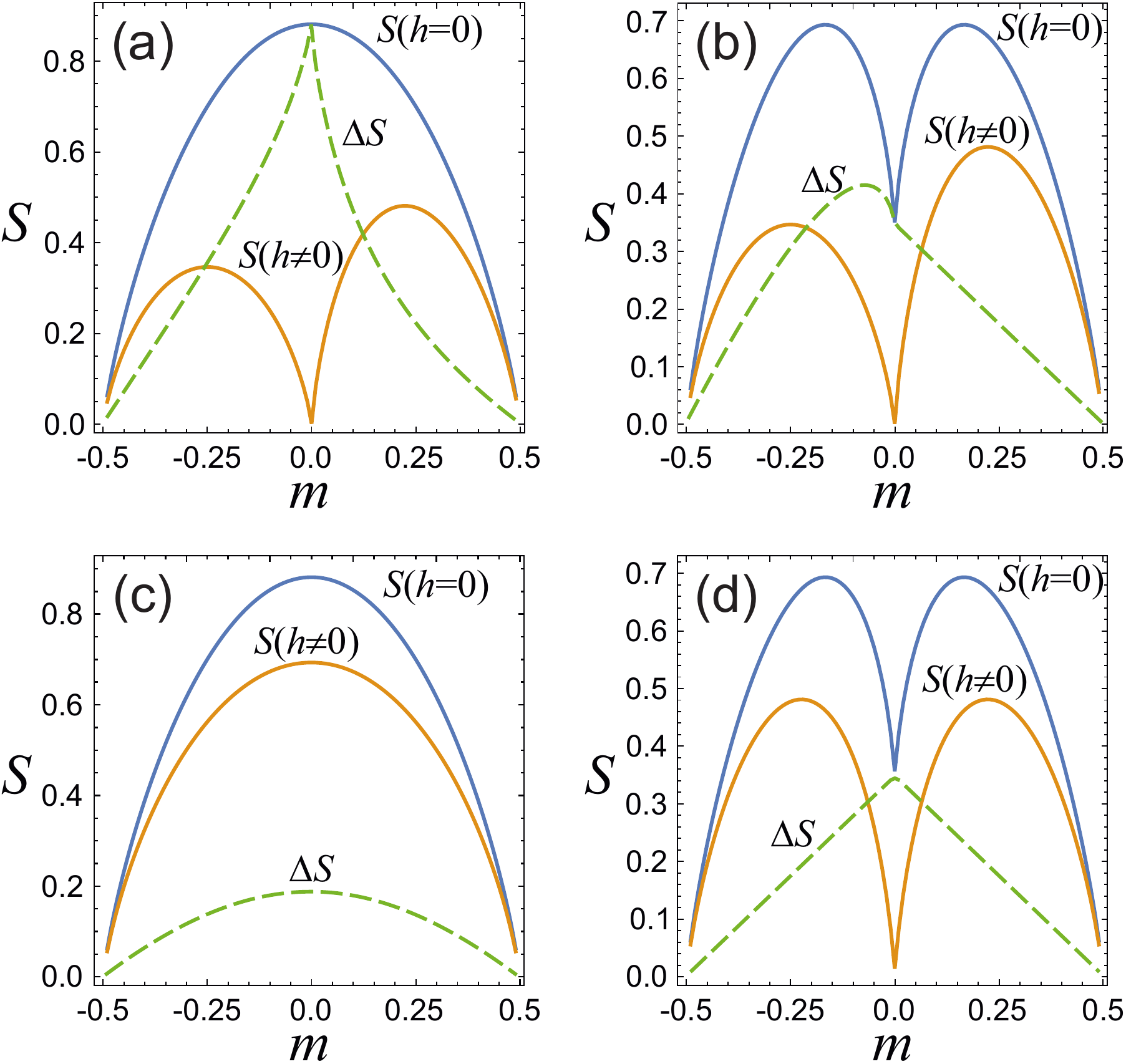}
	\caption{(color online)  
	The dependence of the entropy $S$ (solid lines) and the magnetic entropy change $\Delta S$ (dashed lines) on 
	the deviation of the impurity concentration $n$ from half-filling, $m = n-1/2$, at $T/V=0.05$. 
	Panel (a) corresponds to $J/V=-1$, (b) -- to $J/V=-0.5$, (c) -- to $J/V=1$, and (d) -- to $J/V=0.5$.
	\label{fig:Teq0}
	}
\end{figure}

The concentration dependences of entropy at $T/V=0.05$ shown in Fig.~\ref{fig:Teq0} allow estimating approximately the features of the residual entropy $S_0$. 
The dependences of $S_0$ on $m$ have the following form~\cite{Panov2020}:
\begin{equation}
	S_0 = \ln  \left(  \frac{1}{2} + g_0  \right) 
	+ \frac{1}{2}  \ln \frac{2}{ \frac{1}{4} - m^2 }
	- m  \ln  \frac{ g_0 + m }{ g_0 - m } ,\qquad |J|/V=1 ,
\end{equation}
\begin{equation}
	S_0 = \frac{1}{2}   \ln \frac{ \frac{1}{2} + |m| }{ \frac{1}{2} - |m| } 
	 +  |m|  \ln \frac{ \frac{1}{4} - m^2 }{ 8 m^2 } 
	 + \frac{1}{2}  \ln 2  ,\qquad |J|/V<1 ,
\end{equation}
where
\begin{equation}
	g_0 = \frac{1}{\sqrt{2}} \left( \frac{1}{4} + m^2 \right)^{1/2} .
\end{equation}
These expressions depend only on $|m|$ and are identical for the AFM and FM cases. 
The curves of $S(h=0)$ in Fig.~\ref{fig:Teq0}(a) and (c), and in Fig.~\ref{fig:Teq0}(b) and (d) confirm this property. 
For $h\neq0$ the concentration dependences of the residual entropy become asymmetric with respect to $m=0$ for AFM case, 
but save the symmetry in FM case. 
The same dependence for the AFM and FM cases holds only at $m>0$ for $|J|<V$, when the ground state consists of single spins separated by nonmagnetic impurities, and the sign of the exchange constant has no effect.

\subsection{The isentropic dependence of the temperature on the magnetic field.}

\begin{figure}
\centering
	\includegraphics[width=1\textwidth]{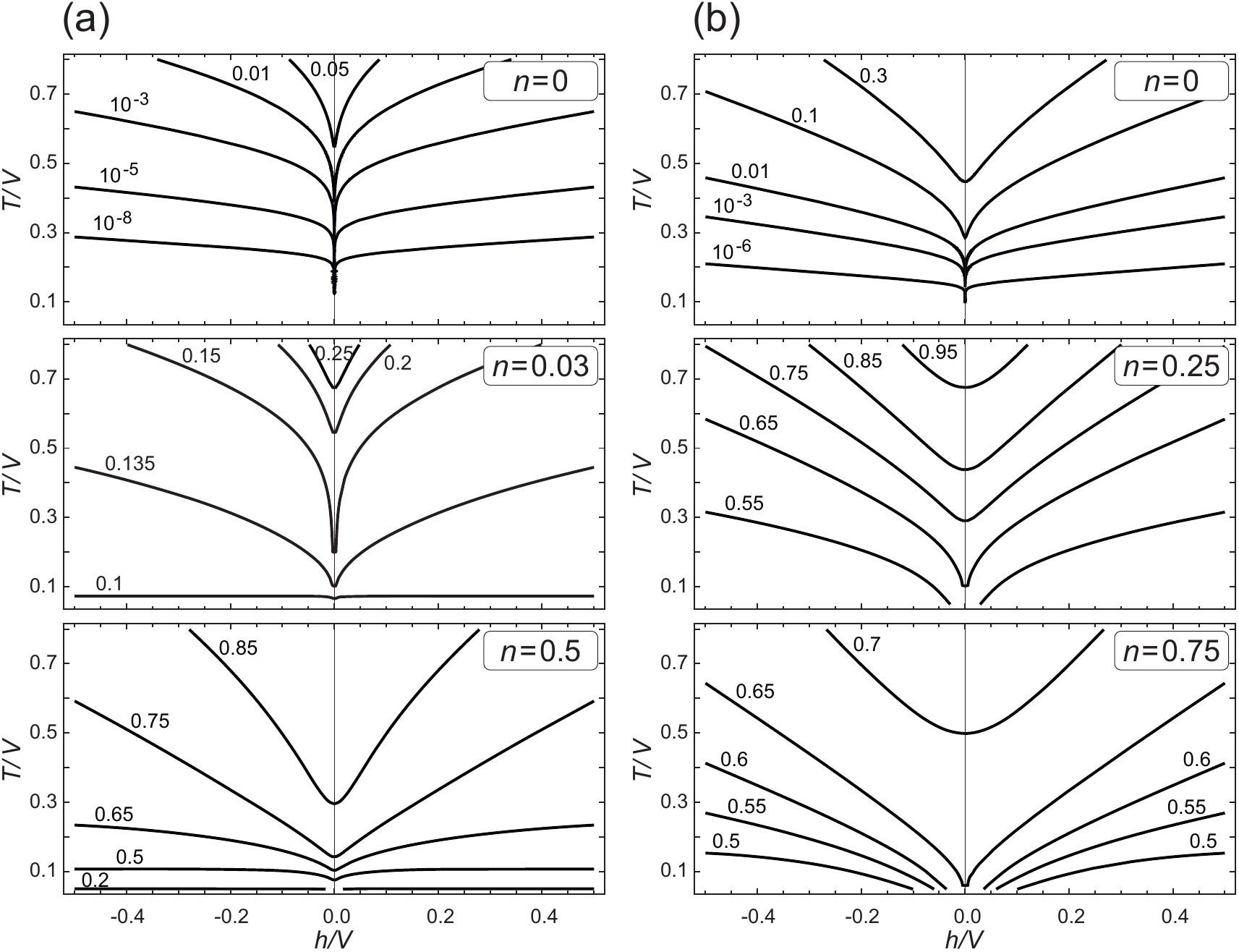}
	\caption{
	The isentropic lines in the $h-T$ parameter plane 
	(a) for $J/V=1.3$ and (b) for $J/V=0.5$.
	The value of the impurity concentration $n$ is given in the frame. 
	The numbers next to the lines show the entropy values. 
	\label{fig:shT-FM}
	}
\end{figure}

Fig.~\ref{fig:shT-FM} shows the isentropic lines in the $h-T$ parameter plane for the ferromagnetic sign of exchange constant, $J>0$. 
For the Ising chain without impurities, the isentropes slope near the critical field $h_c=0$ is almost vertical that leads to extremely high and narrow peak of the Gr\"uneisen parameter~\cite{Gomes2019}, 
which is proportional to 
$e^{2J/T}$ at $h\propto T \,e^{-2J/T}$. 
The impurities change this picture drastically: 
the entropy value increases by several orders of magnitude, 
and the isentropes slope near $h_c=0$ remains finite.

\begin{figure}
\centering
	\includegraphics[width=0.67\textwidth]{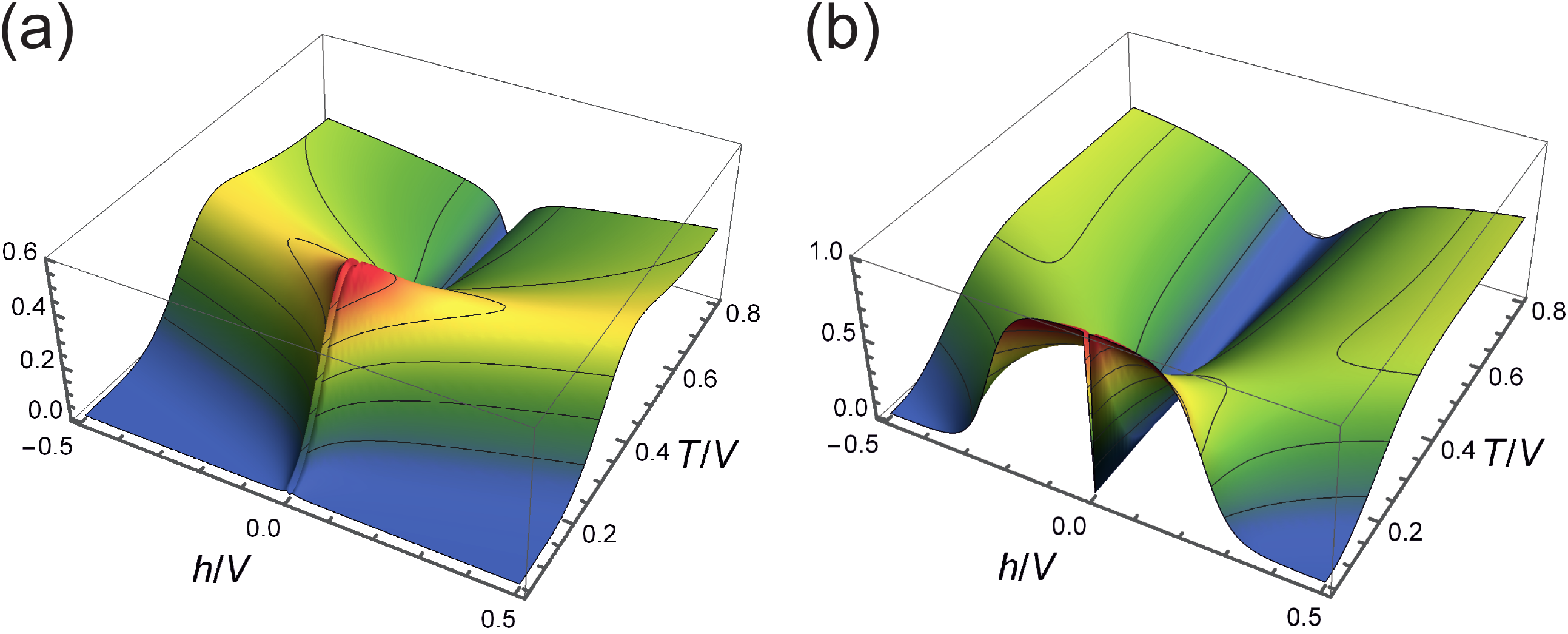}
	\caption{(color online)
	The prefactor of the Gr\"uneisen parameter $-G_r$
	(a) for $J/V=1.3$, $n=0.03$, 
	(b) for $J/V=0.5$, $n=0.25$.
	\label{fig:GH-FM}
	}
\end{figure}

The magnetic Gr\"uneisen parameter can be rewritten~\cite{Wolf2014} in the scaling form as
\begin{equation}
	\Gamma_{mag} = -G_r\frac{1}{h - h_c},
\label{eq:GrunScale}
\end{equation}
where $-G_r$ is a prefactor, and $h_c$ is a critical magnetic field. 
Fig.~\ref{fig:GH-FM} shows the value $-G_r$ for $h_c=0$ as a function of $T$ and $h$ 
for $J/V=1.3$, $n=0.03$ in panel (a), 
and for $J/V=0.5$, $n=0.25$ in panel (b). 
As can be seen, in both cases, impurities lead to suppression of the singular behavior of the magnetic Gr\"uneisen parameter which is observed for the Ising chain without impurities. 
At low temperatures, in the FR-FM state, the system behaves like an ideal paramagnet near $h=0$ 
with a prefactor value $-G_r=1$~\cite{Wolf2014}.

\begin{figure}
\centering
	\includegraphics[width=1\textwidth]{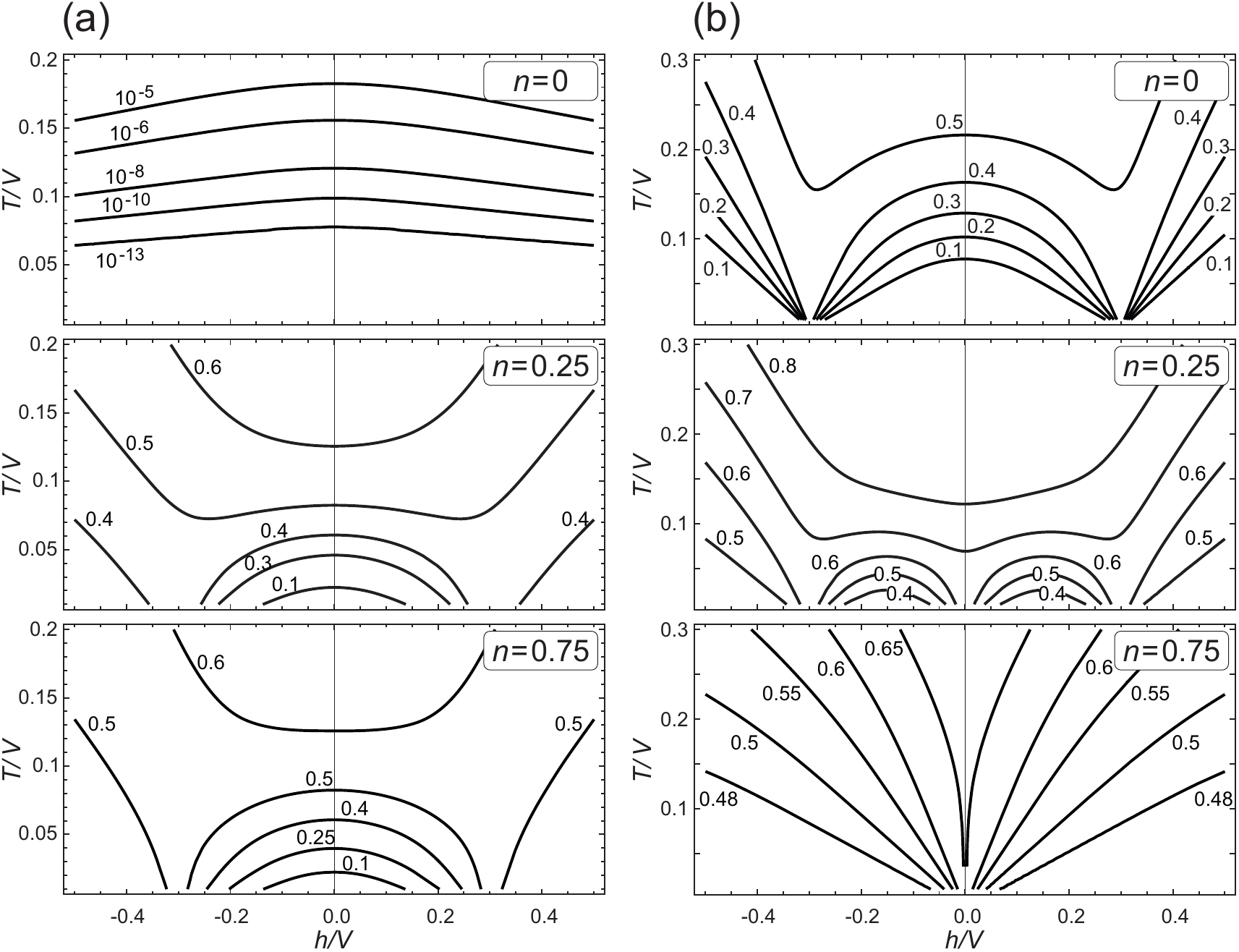}
	\caption{
	The isentropic lines in the $h-T$ parameter plane 
	(a) for $J/V=-1.3$ and (b) for $J/V=-0.15$. 
	The value of the impurity concentration $n$ is given in the frame. 
	The numbers next to the lines show the entropy values. 
	\label{fig:shT-AFM}
	}
\end{figure}

Fig.~\ref{fig:shT-AFM} shows the isentropic lines in the $h-T$ parameter plane for the antiferromagnetic sign of exchange constant, $J<0$. 
The case of a moderate value of the exchange constant, $J/V = -1.3$, is given in panel (a).
In the absence of impurities, there is practically no dependence of entropy on the magnetic field. 
Impurities lead to an increase in the entropy of the system by several orders of magnitude and the appearance of two critical values of the magnetic field, $|h_c|=-J-V$, which correspond to the transition lines from AFM to FR-AFM or FR-PM states. 
It is worth to note, that for comparable values of $|J|$ and $V$, the critical field $|h_c|=-J-V$ can be much smaller than the spin-flip field $|h_c|=-2J$. 
The case of a small exchange constant, $J/V = -0.15$, is given in panel (b). 
Without impurities, the system has two critical spin-flip fields, $|h_c|=-2J$. 
Impurities lead to the appearance of a critical field $h_c=0$. 
As a result, there are three critical fields for a weakly diluted case, 
and only one critical field $h_c=0$ for a strongly diluted case.

\begin{figure}
\centering
	\includegraphics[width=1\textwidth]{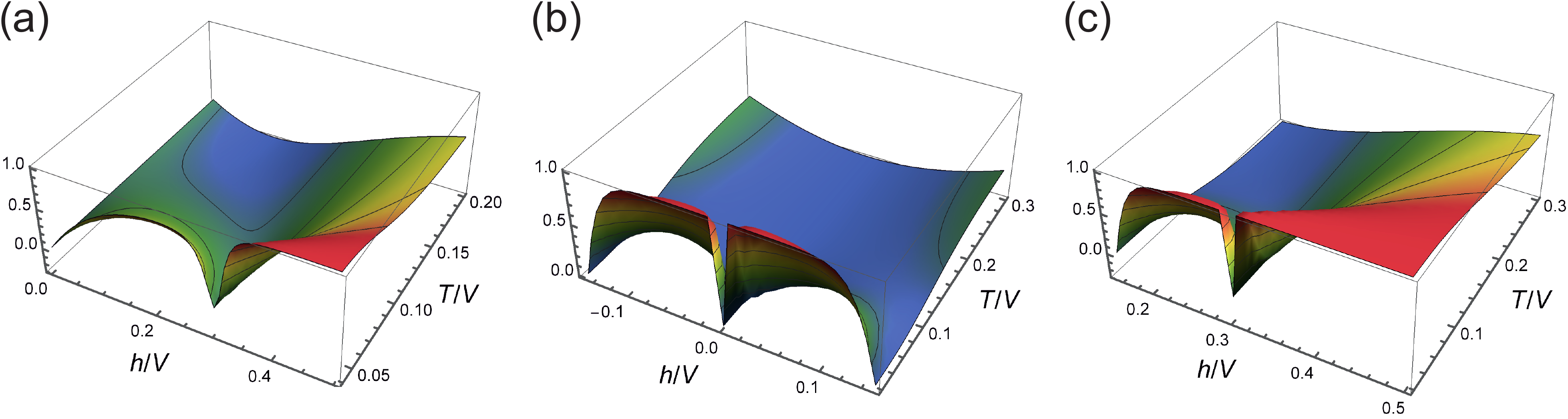}
	\caption{(color online)
	The prefactor of the Gr\"uneisen parameter $-G_r$ near the critical field 
	(a) for $J/V=-1.3$, $n=0.25$, $h_c/V= 0.3$, 
	(b) for $J/V=-0.15$, $n=0.25$, $h_c=0$,
	(c) for $J/V=-0.15$, $n=0.25$, $h_c=0.3$.
	\label{fig:GH-AFM}
	}
\end{figure}

Fig.~\ref{fig:GH-AFM} shows the prefactor of the magnetic Gr\"uneisen parameter 
as a function of $T$ and $h$ near the corresponding critical fields: 
for $J/V=-1.3$, $n=0.25$, $h_c/V= -1 - J/V = 0.3$ in panel (a), 
for $J/V=-0.15$, $n=0.25$, $h_c=0$ in panel (b),
and for $J/V=-0.15$, $n=0.25$, $h_c/V= - 2J/V = 0.3$ in panel (c). 
In all cases, the prefactor tends to $-G_r=1$ at sufficiently low temperatures.

\section{Conclusion}
We examined the effects of the magnetic field on the frustrated phase states of the dilute Ising chain. 
The temperature dependences of entropy and the magnetic entropy change show the nonequivalence of frustrated phases in AFM and FM cases.
The largest effect is achieved when $|J|/V=1$. 
In the AFM case, $J/V=-1$, the nonzero magnetic field causes a charge ordering for nonmagnetic impurities and leads to the maximal value of the magnetic entropy change at a half-filling. 
In the FM case, $J/V=1$, the magnetic field reduces the frustration of the ground state only partially. 
Impurities radically change the magnetic Gr\"uneisen parameter in comparison with the case of a pure Ising chain. 
They suppress the singular behavior of $\Gamma_{mag}$ near $h=0$ in the FM case and produce the paramagnetic behavior in the FR-FM case. 
In the AFM case, additional values of the critical magnetic field for $\Gamma_{mag}$ appear, which are associated with the transition line from AFM to frustrated ground state. 
In the FR-AFM state, $\Gamma_{mag}$ exhibits paramagnetic behavior at $h=0$.

\section*{Acknowledgments}
This work was supported by the Ministry of Education and Science of the Russian Federation, project FEUZ-2020-0054.


\end{document}